\documentclass[11pt]{article}

\usepackage[reqno]{amsmath}
\usepackage{epsfig}
\usepackage{array}
\usepackage{float}

\def\ltap{\ \raisebox{-.4ex}{\rlap{$\sim$}} \raisebox{.4ex}{$<$}\ }
\def\gtap{\ \raisebox{-.4ex}{\rlap{$\sim$}} \raisebox{.4ex}{$>$}\ }
\newcommand{\deltaatm}{\mbox{$\Delta  m^2_{\mathrm{atm}} \ $}}
\newcommand{\deltasol}{\mbox{$ \Delta  m^2_{\odot} \ $}}
\newcommand{\utre}{\mbox{$|U_{\mathrm{e} 3}|$}}
\newcommand{\betabeta}{\mbox{$(\beta \beta)_{0 \nu}  $}}

\newcommand{\meff}{\mbox{$\left|  < \! m  \! > \right| \ $}}
\newcommand{\hbeta}{$\mbox{}^3 {\rm H}$ $\beta$-decay \ }
\newcommand{\eV}{\mbox{$ \  \mathrm{eV} \ $}}
\newcommand{\deltatre}{\mbox{$ \ \Delta m^2_{32} \ $}}
\newcommand{\deltadue}{\mbox{$ \ \Delta m^2_{21} \ $}}

\newcommand{\deltaatmmax}{\mbox{$(\Delta  m^2_{\mathrm{atm}})_{ \! \mbox{}_{\mathrm{MAX}}} \ $}}
\newcommand{\deltaatmmin}{\mbox{$(\Delta  m^2_{\mathrm{atm}})_{ \! \mbox{}_{\mathrm{MIN}}} \ $}}
\newcommand{\deltasolmax}{\mbox{$ (\Delta  m^2_{\odot})_{ \! \mbox{}_{\mathrm{MAX}}} \ $}}
\newcommand{\deltasolmin}{\mbox{$(\Delta  m^2_{\odot})_{ \! \mbox{}_{\mathrm{MIN}}}  \ $}}
\newcommand{\utremax}{\mbox{$|U_{\mathrm{e} 3}|^2_{ \! \mbox{}_{\mathrm{MAX}}}$ }}
\newcommand{\utremin}{\mbox{$|U_{\mathrm{e} 3}|^2_{ \! \mbox{}_{\mathrm{MIN}}} $ }}

\newcommand{\uuno}{\mbox{$|U_{\mathrm{e} 1}|^2$}}
\newcommand{\uunomax}{\mbox{$|U_{\mathrm{e} 1}|^2_{ \! \mbox{}_{\mathrm{MAX}}}$ }}
\newcommand{\uunomin}{\mbox{$|U_{\mathrm{e} 1}|^2_{ \! \mbox{}_{\mathrm{MIN}}}$ }}

\begin{document}

\hfill{Ref. SISSA 33/2002/EP}
\rightline{April 2002}
\rightline{hep-ph/0205022}

\begin{center}
{\bf  The SNO Solar Neutrino Data, Neutrinoless Double Beta-Decay 
and Neutrino Mass Spectrum}

\vspace{0.4cm}

S. Pascoli
~and~
S. T. Petcov 
\footnote{Also at: Institute of Nuclear Research and
Nuclear Energy, Bulgarian Academy of Sciences, 1784 Sofia, Bulgaria}

\vspace{0.2cm}

{\em  Scuola Internazionale Superiore di Studi Avanzati, 
I-34014 Trieste, Italy\\
}
\vspace{0.2cm}   
{\em Istituto Nazionale di Fisica Nucleare, 
Sezione di Trieste, I-34014 Trieste, Italy\\
}
\end{center}

\begin{abstract}

   Assuming 3-$\nu$ mixing and  massive Majorana neutrinos,
we analyze the  implications of the results of the 
solar neutrino experiments, including the latest
SNO data, which favor the LMA MSW solution of the 
solar neutrino problem with
$\tan^2\theta_{\odot} < 1$, 
for the predictions of the
effective Majorana mass \meff in
neutrinoless double beta-decay. 
Neutrino mass spectra with normal mass hierarchy, 
with inverted hierarchy and of quasi-degenerate type 
are considered. For $\cos 2\theta_{\odot} \gtap 0.26$,
which follows (at 99.73\% C.L.)
from the SNO analysis of the solar neutrino data,
we find significant lower limits on \meff
in the cases of quasi-degenerate and 
inverted hierarchy neutrino mass spectrum,
$\meff \gtap 0.035$ eV and                          
$\meff \gtap 8.5\times 10^{-3}$ eV, respectively. If the 
spectrum is hierarchical the upper limit holds 
$\meff   \ltap 8.2\times 10^{-3}$ eV.
Correspondingly, not only a measured value of 
$\meff \neq 0$, but even an experimental
upper limit on \meff of the order of
$\mbox{few} \times 10^{-2}$ eV can provide 
information on the type of
the neutrino mass spectrum; it can provide also 
a significant upper limit on the mass of the lightest
neutrino $m_1$. A measured value of $\meff \gtap 0.2$ eV,
combined with data on neutrino masses
from the $^3$H $\beta-$decay 
experiment KATRIN, 
might allow to establish whether 
the CP-symmetry is
violated in the lepton sector.

\end{abstract}

\newpage
\section{Introduction}
\vspace{-0.1cm}

\hskip 1.0truecm With the publication of the new results 
of the SNO solar neutrino experiment \cite{SNO2,SNO3}
(see also \cite{SNO1}) on
i) the measured rates of the charged current (CC) 
and neutral current (NC) reactions,
$\nu_e + D \rightarrow e^{-} + p + p$
and $\nu_l~(\bar{\nu}_l) + D \rightarrow \nu_l~(\bar{\nu}_l) + n + p$,
ii) on the day-night (D-N) asymmetries  
in the CC and NC reaction rates,
and iii) on the day and night event energy spectra,
further strong evidences 
for oscillations or transitions
of the solar $\nu_e$
into active neutrinos 
$\nu_{\mu,\tau}$ 
(and/or antineutrinos $\bar{\nu}_{\mu,\tau}$),
taking place when the solar $\nu_e$
travel from the central 
region of the Sun to the Earth, 
have been obtained.
The evidences for oscillations (or transitions)
of the solar $\nu_e$ become even stronger 
when the SNO data are combined with the
data obtained in the other solar neutrino experiments,
Homestake, Kamiokande, SAGE, GALLEX/GNO and 
Super-Kamiokande \cite{SKsol,Cl98}.

    Global analysis of the solar neutrino data 
\cite{SNO2,SNO3,SNO1,SKsol,Cl98},
including the latest SNO results,
in terms of the hypothesis of 
oscillations of the solar $\nu_e$ 
into active neutrinos, 
$\nu_e \rightarrow \nu_{\mu (\tau)}$,
show \cite{SNO2} that the data favor the large mixing 
angle (LMA) MSW solution with
$\tan^2\theta_{\odot} < 1$,
where $\theta_{\odot}$ is the 
angle which controls the 
solar neutrino transitions. The LOW 
solution of the 
solar neutrino problem
with transitions into 
active neutrinos is only allowed at
approximately 99.73\% C.L. \cite{SNO2};
there do not exist other solutions
at the indicated confidence level.
In the case of the LMA solution, 
the range of values of the 
neutrino mass-squared difference 
$\Delta m^2_{\odot} > 0$, characterizing
the solar neutrino 
transitions,
found in \cite{SNO2} at 99.73\% C.L. reads:
\begin{equation}
{\rm LMA~MSW}:~~~~~~2.2\times 10^{-5}~{\rm eV^2} 
\ltap \Delta m^2 
\ltap 2.0\times 10^{-4}~{\rm eV^2}~~~~~(99.73\%~{\rm C.L.}).
\label{dmsolLMA}
\end{equation}
%
\noindent The best fit value of $\Delta m^2_{\odot}$
obtained in \cite{SNO2}  
is $(\Delta m^2_{\odot})_{\mathrm{BF}} = 5.0\times 10^{-5}~{\rm eV^2}$.
The mixing angle $\theta_{\odot}$
was found in the case of the LMA solution
to lie in an interval which 
at 99.73\% C.L. is determined by \cite{SNO2}
\begin{equation}
{\rm LMA~~MSW}:~~~~~~~~~~~~~~~~~
0.26 \ltap \cos2\theta_{\odot} \ltap 0.64~~~~~~~(99.73\%~{\rm C.L.}).~~~~~~~~~
\label{thLMA}
\end{equation}
%
\noindent The best fit value of
$\cos2\theta_{\odot}$ 
in the LMA solution region is given by    
$(\cos2\theta_{\odot})_{\mathrm{BF}} = 0.50$.

   Strong evidences for oscillations of 
atmospheric neutrinos have been obtained in the 
Super-Kamiokande experiment \cite{SKatm00}.
As is well known, the atmospheric neutrino data
is best described in terms of
dominant $\nu_{\mu} \rightarrow \nu_{\tau}$
($\bar{\nu}_{\mu} \rightarrow \bar{\nu}_{\tau}$)
oscillations. The explanation of the solar and 
atmospheric neutrino data in terms of 
neutrino oscillations requires
the existence of 3-neutrino mixing
in the weak charged lepton current (see, e.g., \cite{BGG99,P99}).

  Assuming 3-$\nu$ mixing
and massive Majorana neutrinos,
we analyze the implications of the  
latest results 
of the SNO experiment
for the predictions of the
effective Majorana mass \meff in
neutrinoless double beta (\betabeta-) decay 
(see, e.g., \cite{BiPet87,BPP1,PPW}): 
\begin{equation}
\meff = \left| m_1 |U_{\mathrm{e} 1}|^2 
+ m_2 |U_{\mathrm{e} 2}|^2~e^{i\alpha_{21}}
 + m_3 |U_{\mathrm{e} 3}|^2~e^{i\alpha_{31}} \right|.
\label{effmass2}
\end{equation}
\noindent Here $m_{1,2,3}$ are the masses of 3 
Majorana neutrinos with definite mass 
$\nu_{1,2,3}$, 
$U_{{\rm e} j}$ are elements of 
the lepton mixing matrix $U$ -
the Pontecorvo-Maki-Nakagawa-Sakata 
(PMNS) mixing matrix \cite{BPont57,MNS62},
and $\alpha_{21}$ and $\alpha_{31}$ 
are two Majorana CP-violating phases
\footnote{We assume that 
the fields of the 
Majorana neutrinos $\nu_j$ 
satisfy the Majorana condition:
$C(\bar{\nu}_{j})^{T} = \nu_{j},~j=1,2,3$,
where $C$ is the charge conjugation matrix.}
\cite{BHP80,Doi81}.
If CP-invariance holds, 
one has \cite{LW81,BNP84}
$\alpha_{21} = k\pi$, $\alpha_{31} = 
k'\pi$, $k,k'=0,1,2,...$, and 
\begin{equation}
\eta_{21} \equiv e^{i\alpha_{21}} = \pm 1,~~~
\eta_{31} \equiv e^{i\alpha_{31}} = \pm 1 ,
\label{eta2131}
\end{equation}
\noindent represent the relative 
CP-parities of the neutrinos 
$\nu_1$ and $\nu_2$, and 
$\nu_1$ and $\nu_3$, respectively. 

  The experiments searching for \betabeta-decay 
test the underlying symmetries 
of particle interactions (see, e.g., \cite{BiPet87}). They
can answer the fundamental 
question about the nature of 
massive neutrinos, which can be Dirac or Majorana 
fermions. If the massive neutrinos are 
Majorana particles, the observation of \betabeta-decay
\footnote{Evidences for \betabeta-decay
taking place with a rate corresponding to
 $0.11 \ {\rm eV} \leq  \meff \leq  0.56$ eV
(95\% C.L.) are claimed to 
have been obtained in \cite{Klap01}. The
results announced in \cite{Klap01} have been 
criticized in \cite{FSViss02,bb0nu02}.}  
can provide unique information on the type of the 
neutrino mass spectrum and on the lightest neutrino mass
\cite{BPP1,PPW,SPAS94,BGKP96,BGGKP99,bbpapers1,BPP2,bbpapers2}.
Combined with data from the 
\hbeta neutrino mass experiment KATRIN \cite{KATRIN},
it can give also unique information 
on the CP-violation in the lepton sector 
induced by the Majorana CP-violating
phases, and if CP-invariance holds - on the 
relative CP-parities of the massive Majorana 
neutrinos \cite{BPP1,PPW,BGKP96,bbpapers3}. 
   
    Rather stringent upper
bounds on \meff have been obtained in the 
$^{76}$Ge experiments 
by the Heidelberg-Moscow collaboration \cite{76Ge00}, 
$ \meff < 0.35~{\rm eV}$ ($90\%$C.L.), 
and by the IGEX collaboration \cite{IGEX00},
$\meff < (0.33 \div 1.35)~{\rm eV}$ ($90\%$C.L.).
Taking into account a factor of 3 uncertainty
in the calculated value of the corresponding 
nuclear matrix element, we get for the upper limit
found in \cite{76Ge00}:  $\meff < 1.05$ eV.
Considerably higher sensitivity to the value of 
$\meff$ is planned to be 
reached in several $\betabeta-$decay experiments
of a new generation. 
The NEMO3 experiment \cite{NEMO3}, 
which will begin to take data in July of 2002, 
and the cryogenic detector CUORE 
\cite{CUORE}, are expected 
to reach a sensitivity to values of 
$\meff \cong 0.1~$eV.
An order of magnitude better sensitivity,
i.e., to $\meff \cong 10^{-2}~$eV,
is planned to be achieved 
in the GENIUS experiment \cite{GENIUS}  
utilizing one ton of enriched
$^{76}$Ge, and in the 
EXO experiment \cite{EXO}, which will search for
$\betabeta-$decay of $^{136}$Xe.
Two more detectors, Majorana \cite{Maj} 
and MOON \cite{MOON},
are planned to have sensitivity to \meff
in the range of $few \times 10^{-2}$ eV.

In what regards the \hbeta experiments, 
the currently existing most stringent upper 
bounds on the electron (anti-)neutrino mass  
$m_{\bar{\nu}_e}$ were obtained in the
Troitzk~\cite{MoscowH3} and Mainz~\cite{Mainz} 
experiments and read
$m_{\bar{\nu}_e} <  2.2 \eV$.
The KATRIN \hbeta experiment \cite{KATRIN}
is planned to reach a sensitivity  
to  $m_{\bar{\nu}_e} \sim 0.35$ eV.

   The fact that the solar neutrino data implies 
a relatively large lower limit on 
the value of $\cos2\theta_{\odot}$,
eq. (\ref{thLMA}), has important 
implications for the predictions of the 
effective Majorana mass parameter 
in \betabeta-decay \cite{BPP1,PPW}
and in the present article we investigate 
these implications.

\vspace{-0.3cm}
\section{The SNO Data and the 
Predictions for the Effective Majorana Mass \meff}
\vspace{-0.1cm}

\hskip 1.0truecm  According to the analysis 
performed in \cite{SNO2}, the solar neutrino 
data, including the latest SNO results,
strongly favor the LMA solution of the solar 
neutrino problem with $\tan^2 \theta_\odot <1$. 
We take into account these new development to
update the predictions for the effective Majorana mass \meff, 
derived 
in \cite{BPP1},
and the analysis of the implications
of the measurement of, or obtaining a more 
stringent upper limit on,
\meff performed in \cite{BPP1,PPW}. 
The predicted value of
\meff depends in the 
case of 3-neutrino mixing of interest
 on (see e.g. \cite{BPP1,PPW,BPP2}):
i) the value of the lightest neutrino mass $m_1$,
ii) $\Delta m^2_{\odot}$ and $\theta_{\odot}$, 
iii) the neutrino mass-squared difference which
characterizes the atmospheric $\nu_{\mu}$ 
($\bar{\nu}_{\mu}$) oscillations,
$\deltaatm$, and 
iv) the  lepton mixing angle $\theta$ which is
limited by the CHOOZ and Palo Verde 
experiments \cite{CHOOZ,PaloV}.
The ranges of allowed values of 
$\Delta m^2_{\odot}$ and $\theta_{\odot}$
are determined in \cite{SNO2},
while those of $\deltaatm$ and
$\theta$ are taken from \cite{Gonza3nu}
(we use the best fit values and the 
99\% C.L. results from \cite{Gonza3nu}).
Given the indicated parameters,
the value of \meff depends strongly 
\cite{BPP1,PPW} on the type of the
neutrino mass spectrum, as well as 
on the values of the two
Majorana CP-violating phases,
$\alpha_{21}$ and $\alpha_{31}$ 
(see eq. (\ref{effmass2})),
present in the lepton mixing matrix.

  Let us note that if $\deltaatm$ lies in the interval 
$\deltaatm \cong (2.0 - 5.0)\times 10^{-3}~{\rm eV^2}$, 
as is suggested by the current 
atmospheric neutrino data \cite{SKatm00}, 
its value will be determined with a high precision
($\sim 10\%$ uncertainty) by the MINOS experiment
\cite{MINOS}. Similarly, if  
$\Delta m^2_{\odot} \cong (2.5 - 10.0)\times 10^{-5}~{\rm eV^2}$,
which is favored by the solar neutrino data,
the KamLAND experiment will be able to measure
$\Delta m^2_{\odot}$  with an uncertainty of
$\sim (10 - 15)\%$ (99.73\% C.L.). 
Combining the data 
from the solar neutrino experiments 
and from KamLAND would permit to determine
$\cos2\theta_{\odot}$ with a high precision
as well ($\sim 15\%$ uncertainty at 99.73\% C.L., see 
\footnote{We thank C. Pe\~na-Garay for 
clarifications on this point.}
, e.g., \cite{Carlos01}). 
Somewhat better limits on $\sin^2 \theta$ than 
the existing one can be obtained in the MINOS experiment. 

    We number the massive neutrinos 
(without loss of generality)
in such a way that $m_1 < m_2 < m_3$.
In the analysis which follows
we consider neutrino mass 
spectra with normal mass hierarchy, 
with inverted hierarchy and of 
quasi-degenerate type 
\cite{BPP1,PPW,SPAS94,BGKP96,BGGKP99,bbpapers1,bbpapers2}. 
In the case of neutrino mass spectrum with
normal mass hierarchy ($m_1 \ll~(<)~ m_2 \ll m_3$) 
we have $\deltasol \equiv \deltadue$ and 
$\sin^2\theta \equiv |U_{{\rm e} 3}|^2$, 
while in the case of spectrum with
inverted hierarchy ($m_1 \ll  m_2 \cong m_3$)
one finds $\deltasol \equiv \deltatre$
and $\sin^2\theta \equiv |U_{{\rm e} 1}|^2$.
In both cases one can choose
$\deltaatm \equiv \Delta m^2_{31}$.
It should be noted that 
for $m_1 > 0.2  \ {\rm eV} \gg \sqrt{\deltaatm}$,
the neutrino mass spectrum is of the quasi-degenerate type, 
$m_1 \cong  m_2 \cong m_3$, and the two possibilities, 
$\deltasol \equiv \deltadue$
and $\deltasol \equiv \deltatre$, 
lead to the same predictions for \meff. 

\vspace{-0.2cm}
\subsection{Normal Mass Hierarchy: $\deltasol \equiv \deltadue$}

\hskip 1.0truecm  If $\deltasol = \Delta m^2_{21}$,
the effective Majorana mass parameter 
\meff is  given in terms
of the oscillation parameters \deltasol, \deltaatm, 
$\theta_\odot$ and $|U_{\mathrm{e}3}|^2$
which is constrained by the CHOOZ data, as follows \cite{BPP1}:
\begin{equation}
\meff = \left| \big( m_1 \cos^2 \theta_\odot + 
 \sqrt{ m_1^2 + \deltasol} \sin^2 \theta_\odot e^{i\alpha_{21}} \big) 
 (1 - |U_{\mathrm{e}3}|^2) 
+ 
\sqrt{m_1^2 +\deltaatm} |U_{\mathrm{e}3}|^2 e^{i\alpha_{31} } \right|.
\label{eqmasshierarchy01} 
\end{equation}
%
The effective Majorana mass
\meff can lie anywhere between 0 and 
the present upper limits, as Fig. 1 (left panels) shows
\footnote{This statement is valid 
as long as $m_1$ and the CP violating phases which
enter the effective Majorana mass \meff
are not constrained.}.
This conclusion does not change even 
under the most favorable
conditions for the determination of \meff,
namely, even when \deltaatm, \deltasol,
$\theta_{\odot}$ and $\theta$ are known
with negligible uncertainty \cite{PPW}.
Our further conclusions 
for the case of the 
LMA solution of the solar neutrino problem \cite{SNO2}
are illustrated in Fig. 1 (left panels) and 
are summarized below.

\vspace{0.2cm} 
\noindent {\bf Case A: $m_1 < 0.02 \ {\rm eV}$, $m_1 < m_2 \ll m_3$.}
 
  Taking into account the new constraints on
the solar neutrino oscillation parameters following
from the SNO data \cite{SNO2} does not change qualitatively
the conclusions reached in ref.~\cite{BPP1,PPW}.
The maximal value of \meff, $\meff_{ \! \mbox{}_{\rm MAX}}$,
for given $m_1$ reads:
\begin{align}
\meff_{ \! \mbox{}_{\rm MAX}} =  &
\Big( m_1 (\cos^2 \theta_\odot)_{ \! \mbox{}_{\rm MIN}} + 
 \sqrt{ m_1^2 + (\deltasol)_{ \! \mbox{}_{\rm MAX}}}
( \sin^2 \theta_\odot)_{ \! \mbox{}_{\rm MAX}} \Big) 
 (1 - |U_{\mathrm{e}3}|^2_{ \! \mbox{}_{\rm MAX}}) \nonumber \\
                            &+ 
\sqrt{m_1^2 + (\deltaatm)_{ \! \mbox{}_{\rm MAX}}} 
|U_{\mathrm{e}3}|^2_{ \! \mbox{}_{\rm MAX}} ,
\label{meffmaxhierLMA} 
\end{align}
%
\noindent where  $(\cos^2  \!  \! \theta_\odot)_{\mbox{}_{\rm MIN}}$
and $ (\sin^2  \!  \! \theta_\odot)_{\mbox{}_{\rm MAX}}$
are the values corresponding
to $(\tan^2 \theta_\odot)_{\mbox{}_{\rm MAX}}$,
and $\deltaatmmax$ is the maximal value of  
\deltaatm allowed for the  $\utremax$ \cite{Gonza3nu}.

    For the values of \deltasol and $\tan^2\theta_{\odot}$
from the LMA solution region \cite{SNO2}, 
eqs. (\ref{dmsolLMA}) and (\ref{thLMA}),  
we get for $m_1 \ll 0.02 \ {\rm eV}$:
$\meff_{ \! \mbox{}_{\rm MAX}}
\simeq 8.2 \times 10^{-3}~{\rm eV}$.
Using the best fit values of the oscillation parameters
found in refs.~\cite{SNO2,Gonza3nu}, one obtains:
$\meff_{ \! \mbox{}_{\rm MAX}}
\simeq 2.0 \times 10^{-3}~{\rm eV}$.
The maximal value of \meff corresponds to
the case of CP-conservation and $\nu_{1,2,3}$
having identical CP-parities, $\eta_{21} = \eta_{31}=1$.

   There is no significant lower bound 
on \meff because of the possibility of 
mutual compensations
between the terms contributing
to \meff and corresponding to the exchange of different 
virtual massive Majorana neutrinos.
Furthermore, the uncertainties in the oscillation parameters
do not allow to identify a ``just-CP violation''
region of values of \meff ~\cite{BPP1} 
(a value of \meff in this
region would unambiguously signal the existence of
CP-violation in the lepton sector, caused by 
Majorana CP-violating phases).  
However, if the neutrinoless 
double beta-decay will be observed, 
the measured value of
\meff, combined with information
on $m_1$ and  a better determination
of the relevant neutrino 
oscillation parameters,
might allow to determine
whether the CP-symmetry is violated 
due to Majorana 
CP-violating phases, or to identify which are
the allowed patterns of the massive neutrino CP-parities 
in the case of CP-conservation
(for a detailed discussion see ref.~\cite{PPW}).
 
\vspace{0.2cm} 
\noindent {\bf Case B: 
Neutrino Mass Spectrum with Partial Hierarchy 
($0.02 \ {\rm eV} \leq  m_1 \leq 0.2 \ {\rm eV}$)} 

  For   $m_1 \geq 0.02 \ {\rm eV}$
there exists a 
lower bound on the possible
values of \meff (Fig. 1, left panels). 
Using the 99.73\% C.L. allowed values 
of \deltasol and $\cos2\theta_{\odot}$
from \cite{SNO2}, we find that
this lower bound is  significant,
i.e., $\meff \gtap 10^{-2}$ eV, for $m_1 \gtap 0.07$ eV.
For the best fit values of the oscillation parameters
obtained in \cite{SNO2,Gonza3nu}, one has
$\meff \gtap 10^{-2}$ eV for $m_1 \geq 0.02$ eV.

 For a fixed $m_1 \geq 0.02 \ {\rm eV}$, 
the minimal value of \meff, $\meff_{ \! \mbox{}_{\rm MIN} }$, 
is given by
\begin{equation}
\label{minmeffLMA}
  \meff_{ \! \mbox{}_{\rm MIN} } \! \cong   
  m_1 (\cos 2  \theta_\odot )_{\! \mbox{}_{\rm MIN}} 
\!  (1 - \utremax  \! \! ) -
\sqrt{m_1^2 + \deltaatmmax} \utremax + 
{\cal O} \Big(\frac{\deltasolmax}{4 m_1} \Big) ,
\end{equation}
%
where again $\deltaatmmax$ is the maximal 
allowed value of \deltaatm for the $\utremax$ 
\cite{Gonza3nu}.
   The upper bound on \meff, which corresponds 
to CP-conservation and $\eta_{21} = \eta_{31} = +1$
($\nu_{1,2,3}$
possessing identical CP-parities), can be found
for given $m_1$ by using 
eq.~(\ref{meffmaxhierLMA}).
For the allowed values of $m_1$, 
$ 0.02 \ {\rm eV} \leq m_1 \leq 0.2 \ {\rm eV}$,
we have 
 $\meff \leq 0.2~{\rm eV}$. 

\vspace{-0.2cm}
\subsection{Inverted Neutrino Mass Hierarchy: $\deltasol \equiv \deltatre$}
 
\hskip 1.0cm  If $\deltasol = \Delta m^2_{32}$,
the effective Majorana mass \meff is given in terms
of the oscillation parameters \deltasol, \deltaatm, 
$\theta_\odot$ and 
$|U_{\mathrm{e}1}|^2$
which is constrained by the CHOOZ data \cite{BPP1}:
\begin{align}
\meff \! =  & \Big| m_1 |U_{\mathrm{e}1}|^2  
  + 
 \sqrt{m_1^2 + \deltaatm \! \!  - \deltasol}
 \cos^2 \theta_\odot (1 - |U_{\mathrm{e}1}|^2)  e^{i\alpha_{21}}
 \nonumber \\ 
& + 
 \sqrt{ m_1^2 + \deltaatm} \sin^2 \theta_\odot 
 (1 - |U_{\mathrm{e}1}|^2) e^{i\alpha_{31} }
 \Big|.
\label{eqmassinvhierarchy01} 
\end{align}
%
The new predictions for \meff differ 
substantially from those
obtained before the appearance of the latest
SNO data
due to the existence 
of a significant lower bound on \meff
for every value of $m_1$:
even in the case of $m_1 \ll m_2 \cong m_3$
(i.e., even if $m_1 \ll 0.02$ eV),
we get 
\begin{equation}
\meff \gtap  8.5 \times 10^{-3}~{\rm eV}
\label{meffminIMH}
\end{equation}
(see Fig. 1, right panels).
Given the neutrino oscillation parameters,
the minimal allowed value of \meff
depends on the values of the CP violating phases 
$\alpha_{21}$ and $\alpha_{31}$.

\vspace{0.2cm}
\noindent {\bf Case A: $m_1 < 0.02 \ {\rm eV}$, 
$m_1 \ll m_2 \simeq m_3$. }

\hskip 1.0cm  The effective Majorana mass \meff 
can be considerably larger
than in the case of a 
hierarchical neutrino mass spectrum
\cite{BPP1,BGGKP99}.
The maximal value of \meff corresponds to
CP-conservation and 
$\eta_{21} = \eta_{31}= +  1$,
and for given $m_1$ reads:
\begin{align}
\meff_{\! \mbox{}_{\rm MAX}}
 =  & \ m_1 \uunomin +
 \Big( \sqrt{m_1^2 + \deltaatmmax \! \!  - \deltasolmin}
 (\cos^2   \theta_\odot)_{\! \mbox{}_{\rm MIN}}  \nonumber \\ 
 & + \sqrt{m_1^2 + \deltaatmmax} 
(\sin^2   \theta_\odot)_{\! \mbox{}_{\rm MAX}}\Big)
 (1 - \uunomin)  ,
\label{meffinvmin} 
\end{align}
%
where  $(\cos^2  \!  \! \theta_\odot)_{\mbox{}_{\rm MIN}}$
and $ (\sin^2  \!  \! \theta_\odot)_{\mbox{}_{\rm MAX}}$
are the values corresponding
to $(\tan^2 \theta_\odot)_{\mbox{}_{\rm MAX}}$,
and $\uunomin$ is the minimal allowed value
of $\uuno$ for the $\deltaatmmax$.
For the allowed ranges - 
eqs. (1) and (2) for \deltasol 
and $\tan^2\theta_{\odot}$, and the best fit values 
of the neutrino oscillation
parameters, found in \cite{SNO2,Gonza3nu},
we get
$|<m>|_{\mbox{}_{\rm MAX}} \simeq 0.080 \ {\rm eV}$ and
$|<m>|_{\mbox{}_{\rm MAX}} \simeq 0.056 \ {\rm eV}$, 
respectively.

    There exists a non-trivial  
lower bound on \meff in the case of the LMA solution
for which $\cos 2 \theta_\odot$ is found to be
significantly different from zero.
For the 99.73\%~C.L. allowed values 
of \deltasol and $\cos 2 \theta_\odot$ \cite{SNO2},
this lower bound reads: $\meff \gtap 8.5 \times 10^{-3}~{\rm eV}$.
Using the best fit values of the oscillation parameters 
\cite{SNO2,Gonza3nu}, we find:
\begin{equation}
\meff \gtap 2.8 \times 10^{-2}~{\rm eV}.
\label{bfmeffmin}
\end{equation}
The lower bound is present even for
$\cos 2 \theta_\odot > 0.1$: in this case
$\meff \gtap 4 \times 10^{-3}~{\rm eV}$.
The  minimal value 
of \meff, $\meff_{\! \mbox{}_{\rm MIN}}$,
is reached in the case of CP-invariance and
$\eta_{21} = - \eta_{31} = - 1$, and  
is determined by:
\begin{align}
\meff_{\! \mbox{}_{\rm MIN}}
 =  & \Big| \ m_1 \uunomax -
 \Big( \sqrt{m_1^2 + \deltaatmmin \! \!  - \deltasolmax}
 (\cos^2   \theta_\odot)_{\! \mbox{}_{\rm MIN}}  \nonumber \\ 
 & - \sqrt{m_1^2 + \deltaatmmin} (\sin^2   \theta_\odot)_{\! \mbox{}_{\rm MAX}}
 \Big) (1 - \uunomax)  \ \Big|,
\label{meffinvmina} 
\end{align}
%
where  $(\cos^2  \!  \! \theta_\odot)_{\mbox{}_{\rm MIN}}$
and $ (\sin^2  \!  \! \theta_\odot)_{\mbox{}_{\rm MAX}}$
are the values corresponding
to $(\tan^2 \theta_\odot)_{\mbox{}_{\rm MAX}}$,
and \uunomax is the maximal allowed value of 
$\uuno \ $ for the  \deltaatmmin.
In the two other CP conserving cases of 
$\eta_{21} =  \eta_{31} = \pm 1$,
the lower bound on \meff depends
weakly on the allowed values of 
$\theta_\odot$ and reads 
$\meff \gtap 0.03~{\rm eV}$.

 If the neutrino mass spectrum
is of the inverted hierarchy type,
a sufficiently precise determination
of \deltaatm, $\theta_\odot$
and \uuno (or a better upper limit on
\uuno),  combined with a measurement of \meff
in the current or future \betabeta-decay
experiments,
could allow one to get information
on the difference
of the Majorana CP-violating phases
$(\alpha_{31} - \alpha_{21})$
\cite{BGKP96}.
The value of $\sin^2 (\alpha_{31} - \alpha_{21})/2$
is related to the experimentally
measurable quantities as follows \cite{BPP1,BGKP96}:
%
\begin{equation}
\sin^2 \frac{\alpha_{31} \! - \! \alpha_{21}}{2}
\simeq \Big( 1 - \frac{\meff^2}{(m_1^2 + \deltaatm \!) 
(1 - \uuno)^2} \Big) \frac{1}{\sin^2 2  \theta_\odot}
\simeq \Big( 1 - \frac{\meff^2}{\deltaatm \!
(1 - \uuno)^2} \Big) \frac{1}{\sin^2 2  \theta_\odot},
\label{alpha2131}
\end{equation}
%
($m_1 < 0.02 \ {\rm eV}$). The constraints on
$\sin^2 (\alpha_{31} - \alpha_{21})/2$
one could derive on the basis of eq. (\ref{alpha2131})
are illustrated 
\footnote{Note that the 
CP-violating phase
$\alpha_{21}$ is not constrained
in the case under discussion.
Even if it is found that 
$\alpha_{31} - \alpha_{21}= 0, \pm \pi$,
$\alpha_{21}$ can be a source of 
CP-violation in 
$\Delta L =2$ processes other than
\betabeta-decay.}
in Fig. 11 of ref. \cite{BPP1}. 
Obtaining an experimental upper limit on \meff
of the order of 
0.03  eV would permit,
in particular,
to get a lower bound on the value of
$\sin^2 (\alpha_{31} - \alpha_{21})/2$
and possibly exclude the CP conserving case 
corresponding to $\alpha_{31} - \alpha_{21}= 0$
(i.e., $\eta_{21}= \eta_{31}= \pm 1$). 

\vspace{0.2cm}
\noindent {\bf Case B: 
Spectrum with Partial Inverted Hierarchy
($0.02 \ {\rm eV} \leq m_1 \leq 0.2 \ {\rm eV}$).}

\hskip 1.0cm  The discussion and conclusions 
in the case of the spectrum
with partial inverted hierarchy are identical 
to those in the same case for the 
neutrino mass spectrum with normal hierarchy
given in sub-section 2.1, Case B,
except for the maximal and minimal values of \meff,
$\meff_{ \! \mbox{}_{\rm MAX}}$
and $\meff_{ \! \mbox{}_{\rm MIN}}$,
which for a fixed $m_1$ are determined by:
\begin{align}
\label{maxmeffLMAinva}
  \meff_{ \! \mbox{}_{\rm MAX}} \! \simeq &   \
     m_1    \uunomin   +
\sqrt{m_1^2 + \deltaatmmax}(1 - \uunomin   \! )  , \\
 \label{maxmeffLMAinvb}
 \meff_{ \! \mbox{}_{\rm MIN} } \! \simeq &  
 \left|  m_1    \uunomax - \sqrt{m_1^2 + \deltaatmmin}
(\cos 2  \theta_\odot )_{\! \mbox{}_{\rm MIN}} \! (1 - \uunomax  \! \! )
\right|,
\end{align}
%
$\uunomin $ ($\uunomax$) in eq.~(\ref{maxmeffLMAinva}) 
(in eq.~(\ref{maxmeffLMAinvb})) being the minimal 
(maximal) allowed value
of $\uuno$ given the maximal (minimal) value $\deltaatmmax$ 
($\deltaatmmin$).

   For any $m_1 \geq 0.02 \ {\rm eV}$, the lower bound on \meff
reads: $\meff \gtap 0.01 \ {\rm eV}$. 
Using the best fit values of the neutrino 
oscillation parameters, obtained 
in \cite{SNO2,Gonza3nu}, one finds: $\meff \gtap 0.03 \ {\rm eV}$.

\vspace{-0.2cm}
\subsection{Quasi-Degenerate
Mass Spectrum ($m_1 > 0.2 \ {\rm eV}$, $m_1 \simeq m_2 \simeq m_3 
\simeq m_{\bar{\nu}_e}$)}


\hskip 1.0cm The new element in the predictions for 
\meff in the case of  quasi-degenerate neutrino mass spectrum, 
$m_1 > 0.2 \ {\rm eV}$, is   
the existence of a lower bound on the possible
values of \meff (Fig. 1). 
The lower limit on \meff is reached 
in the case of CP-conservation 
and $\eta_{21}=  \eta_{31}= - 1$.
One finds a  significant lower limit, 
$\meff \gtap 0.01 \ {\rm eV}$, if
\begin{equation}
(\cos 2  \theta_\odot )_{\! \mbox{}_{\rm MIN}} >
{\rm max} \left (0.05, 1.5 
\, \utremax / (1 - \utremax  \! \! ) \right). 
\end{equation}
More specifically, using the best fit value, 
and the $90 \%$ C.L. and the $99.73 \%$ C.L. allowed values,
of $\cos2\theta_{\odot}$ from \cite{SNO2},
we obtain, respectively: $\meff \geq 0.10  \ {\rm eV}$,
$\meff \geq 0.06  \ {\rm eV}$ and 
\begin{equation}
\meff \geq 0.035~{\rm eV}
\label{meffminQDS}
\end{equation}
%
(Figs. 1 and 3).
These values of $ \meff $
are in the range of sensitivity
of the current and future 
\betabeta-decay experiments.

 The upper bound on \meff, which corresponds 
to CP-conservation and $\eta_{21} = \eta_{31} = +1 $
($\nu_{1,2,3}$ 
possessing the same CP-parities), can be found
for a given $m_1$ by using 
eq.~(\ref{meffmaxhierLMA}).
For the allowed values of 
$m_1 > 0.2 \ {\rm eV}$ 
(which is limited from above by 
the $^{3}$H $\beta-$decay data
\cite{MoscowH3,Mainz},
$m_{1,2,3} \simeq m_{\bar{\nu}_e}$),
$\meff_{ \! \mbox{}_{\rm MAX}}$
is limited by the upper bounds 
obtained in the \betabeta-decay 
experiments: 
$\meff < 0.35~{\rm eV}$~\cite{76Ge00}
and $\meff < ( 0.33 - 1.35)~{\rm eV}$~\cite{IGEX00}.

    In the case of CP conservation and 
$\eta_{21}= \pm \eta_{31}= + 1$,
\meff is constrained to lie in the interval \cite{BPP1}
$m_{\bar{\nu}_\mathrm{e}} ( 1 - 2 \utremax)  \leq 
\meff \leq m_{\bar{\nu}_\mathrm{e}}$.
An upper limit on \meff
would lead to an upper limit
on $m_{\bar{\nu}_\mathrm{e}}$ which 
is more stringent than the one obtained in the present
\hbeta experiments: for $\meff < 0.35 \ (1.05) \ {\rm eV}$
we have $m_{\bar{\nu}_\mathrm{e}} < 0.41 \  (1.23) \ {\rm eV}$.
Furthermore, the upper limit 
$\meff < 0.2\ {\rm eV}$
would permit to exclude the CP-parity
pattern $\eta_{21}= \pm \eta_{31}= + 1$
for the quasi-degenerate neutrino mass
spectrum. 

   If the CP-symmetry holds and 
$\eta_{21}= \pm \eta_{31}= - 1$,
there are both an upper and a lower limits on \meff,
$m_{\bar{\nu}_\mathrm{e}} 
((\cos 2  \theta_\odot )_{\! \mbox{}_{\rm MIN}} ( 1 - \utremin)
+ \utremin ) \leq \meff
\leq m_{\bar{\nu}_\mathrm{e}}
((\cos 2  \theta_\odot )_{\! \mbox{}_{\rm MAX}} ( 1 - \utremax)
+ \utremax )$.
Using eq.~(\ref{thLMA}) and
the results on $\utre^2$ from ref.~\cite{Gonza3nu},
one finds $ 0.26 \  m_{\bar{\nu}_\mathrm{e}}  \leq \meff
\leq 0.67 \  m_{\bar{\nu}_\mathrm{e}}$.
Given the allowed values of $\cos2\theta_{\odot}$,
eq. (\ref{thLMA}), the observation of the
\betabeta-decay in 
the present and/or future 
\betabeta-decay experiments,
combined with a sufficiently 
stringent upper bound on 
$m_{\bar{\nu}_e} \simeq m_{1,2,3}$
from the tritium beta-decay experiments,
$m_{\bar{\nu}_e} < 
\meff_{exp} / ((\cos 2  \theta_\odot )_{\! \mbox{}_{\rm MAX}}
(1 - \utremax  \! \! ) + \utremax)$,
would allow one, in particular, to exclude
the case of CP-conservation with 
$\eta_{21}=\pm \eta_{31}= - 1$ (Fig. 2).

  For values of \meff, which are in the range 
of sensitivity of the future \betabeta-decay 
experiments, there exists a ``just-CP-violation''
region. This is illustrated in
Fig. 2, where we show $\meff / m_1$
for the case of quasi-degenerate neutrino mass spectrum,
$m_1 > 0.2 \ {\rm eV}$, $m_1 \simeq m_2 \simeq m_3 \simeq m_{\bar{\nu}_e}$,
as a function of $\cos 2 \theta_\odot$.
The ``just-CP-violation'' interval of values of
$\meff/m_1$ is determined by
\begin{equation}
(\cos 2  \theta_\odot )_{\! \mbox{}_{\rm MAX}}
(1 - \utremax  \! \! ) + \utremax < 
\frac{\meff}{m_{\bar{\nu}_e}} < 1 - 2 \utremax.
\label{CPviol}
\end{equation}
%
Taking into account eq. (\ref{thLMA}) and the
existing limits on $|U_{{\rm e} 3}|^2$, 
this gives $0.67 < \meff/m_{\bar{\nu}_e} < 0.85$.
Information about the masses $m_{1,2,3} \cong m_{\bar{\nu}_e}$ 
can be obtained in the KATRIN experiment \cite{KATRIN}.

  A rather precise determination of \meff, 
$m_1 \cong m_{\bar{\nu}_e}$,
$\theta_{\odot}$ and $|U_{\mathrm{e}3}|^2$ 
would 
imply an interdependent
constraint on the two CP-violating phases
$\alpha_{21}$ and $\alpha_{31}$~\cite{BPP1}
(see Fig. 16 in \cite{BPP1}).
For $m_1 \equiv m_{\bar{\nu}_e} > 0.2 $ eV,
the CP-violating phase $\alpha_{21}$
could be tightly constrained if 
$\utre^2$ is sufficiently small and 
the term in \meff containing 
it can be neglected, as is suggested by the current 
limits on $\utre^2$: 
%
\begin{equation}
\sin^2 \frac{\alpha_{21}}{2} \simeq 
 \Big( 1 - \frac{\meff^2}{m_{\bar{\nu}_e}^2} \Big)
 \frac{1}{\sin^2 2  \theta_\odot}. 
\label{CPviol1}
\end{equation}
%
The term  which depends 
on the CP-violating phase $\alpha_{31}$ 
in the expression for \meff,
is suppressed  by the factor $\utre^2$. Therefore 
the constraint one could possibly obtain on 
$\cos \alpha_{31}$ is trivial
(Fig. 16 in \cite{BPP1}), unless
$ \utre^2 \sim {\cal O} ( \sin^2 \theta_\odot)$.

\vspace{-0.3cm}
\section{The Effective Majorana Mass 
and the Determination of the Neutrino Mass Spectrum}
\vspace{-0.1cm}

\hskip 1.0cm The existence of a lower bound on \meff
in the cases of inverted mass hierarchy 
($\deltasol = \Delta m^2_{32}$)
and quasi-degenerate
neutrino mass spectrum,
eqs. (\ref{meffminIMH}) and (\ref{meffminQDS}), 
implies that the future \betabeta-decay
experiments might allow to 
determine the type of the neutrino
mass spectrum 
(under the general assumptions of 
3-neutrino mixing and massive Majorana neutrinos, 
\betabeta-decay generated 
only by the (V-A) charged current weak interaction 
via the exchange of the three Majorana neutrinos,
neutrino oscillation explanation of the solar 
and atmospheric neutrino data).
This conclusion is valid
not only under the assumption that
the \betabeta-decay will be observed in
these experiments and \meff will be measured, 
but also in the case only a sufficiently stringent upper 
limit on \meff will be derived.

   More specifically, as is illustrated in Fig. 3,
the following statements can be made:
\begin{enumerate}
\item
a measurement of $\meff = \meff_{exp} >
 0.20 \ {\rm eV}$,
would imply that the neutrino mass spectrum
is of the quasi-degenerate type 
($m_1 > 0.20 \ {\rm eV}$) and
that there are both a lower and an upper limit
on $m_1$, $(m_1)_{min} \leq m_1 \leq (m_1)_{max}$.
The values of $(m_1)_{max}$ and $(m_1)_{min}$
are fixed respectively by the equalities
$\meff_{\mbox{}_{\rm MIN}} =  \meff_{ \! exp}$ and
$\meff_{\mbox{}_{\rm MAX}} =  \meff_{ \! exp}$,
where $\meff_{\mbox{}_{\rm MIN}}$ and
$\meff_{\mbox{}_{\rm MAX}}$
are given by eqs.~(\ref{minmeffLMA}) and 
~(\ref{meffmaxhierLMA});

\item
if \meff is measured and is found to lie in the interval
 $8.5 \times 10^{-2} \ {\rm eV}\ltap 
\meff_{exp} \ltap 0.20 \ {\rm eV}$,
one could conclude that
either

i) $\deltasol \equiv \deltadue$ and the spectrum is
of the  quasi-degenerate type 
($m_1 > 0.20  \ {\rm eV}$)
or with partial hierarchy 
($0.02 \ {\rm eV} \leq m_1 \leq 0.2 \ {\rm eV}$),
with $ 8.4 \times 10^{-2} \ {\rm eV} \ltap  m_1 
\ltap  1.2\ {\rm eV}$,
where the maximal and minimal values of $m_1$ are
determined as in the {\it Case 1};

or that 
ii) $\deltasol \equiv \deltatre$ and the spectrum
is quasi-degenerate  
($m_1 > 0.20 \  {\rm eV}$) or with
partial inverted hierarchy 
($ 0.02 \ {\rm eV} \leq m_1 \leq 0.2 \ {\rm eV}$),
with $(m_1)_{min} =  2.0 \times 10^{-2} \ {\rm eV}$
and $ (m_1)_{max} =   1.2  \ {\rm eV}$,
where $(m_1)_{max}$ and $(m_1)_{min}$ 
are given by the equalities
$\meff_{\mbox{}_{\rm MIN}} =  \meff_{ \! exp}$
and $\meff_{\mbox{}_{\rm MAX}} =  \meff_{ \! exp}$,
and $\meff_{\mbox{}_{\rm MIN}}$
and $\meff_{\mbox{}_{\rm MAX}}$
are determined by eqs.~(\ref{maxmeffLMAinvb})
and (\ref{maxmeffLMAinva});

\item
a measured value of \meff satisfying
$8.5 \times 10^{-3} \ {\rm eV}\ltap 
 \meff_{exp} \ltap 8.0 \times 10^{-2} \ {\rm eV}$,
would imply that (see Fig. 3) either 

i) $\deltasol \equiv \deltadue$ 
and the spectrum is of quasi-degenerate type
($m_1 > 0.20 \  {\rm eV}$),
with $(m_1)_{max} \ltap 0.48 \ {\rm eV}$,
or with partial hierarchy 
($0.02 \ {\rm eV} \leq m_1 \leq 0.2 \ {\rm eV}$),

or that ii)  $\deltasol \equiv \deltatre$ and the spectrum
is quasi-degenerate
($m_1 > 0.20 \  {\rm eV}$),
or with partial inverted hierarchy 
($0.02 \ {\rm eV} \leq m_1 \leq 0.2 \ {\rm eV}$),
or with inverted hierarchy ($m_1 < 0.02 \ {\rm eV}$),
with only a significant upper bound on $m_1$, 
$ (m_1)_{min} = 0$,
$(m_1)_{max}  \ltap 0.48 \ {\rm eV}$,
where $(m_1)_{max}$ is determined by 
the equation
$\meff_{\mbox{}_{\rm MIN}} =  \meff_{ \! exp}$,
with $\meff_{\mbox{}_{\rm MIN}}$ given
by eq.~(\ref{maxmeffLMAinvb});

\item
a measurement or an upper limit on \meff,
$ \meff  \ltap 8.0 \times 10^{-3} \ {\rm eV}$,
would lead to the conclusion that 
the neutrino mass spectrum is
of the normal mass hierarchy type,
$\deltasol \equiv \deltadue$, and 
that $m_1$ is limited from above  by
$ m_1  \leq (m_1)_{max} \simeq 5.8 \times 10^{-2}   \ {\rm eV}$,
where $(m_1)_{max}$ is determined by 
the condition $\meff_{\mbox{}_{\rm MIN}} =  \meff_{ \! exp}$,
with $\meff_{\mbox{}_{\rm MIN}}$ given
by eq.~(\ref{minmeffLMA}). 
For the allowed values of the oscillation parameters
(at a given confidence level, Fig. 3),
an upper bound on \meff,  
$\meff < 8 \times 10^{-4} \ {\rm eV}$,
would imply an upper limit
on $m_1$, $m_1 < 0.01 \ {\rm eV}$ - Fig.~3, middle panel,
and $m_1 < 0.025 \ {\rm eV}$ - Fig.~3, lower panel.
For the best fit values of 
\deltaatm, \deltasol, $\theta_\odot$ and $\theta$,
the bound $\meff  \ltap 8 \times 10^{-4} \ {\rm eV}$
would lead to a rather narrow interval of possible values
of $m_1$, $ 1 \times 10^{-3}   \ {\rm eV}
< m_1 <  4 \times 10^{-3}   \ {\rm eV}$ (Fig.~3, upper panel).
\end{enumerate}

 Thus, a measured value of (or an upper limit on)
the effective Majorana mass  $\meff \ltap 0.03  \ {\rm eV}$
would disfavor (if not rule out) the 
quasi-degenerate mass spectrum,
while a value of $\meff \ltap 8 \times 10^{-3} \ {\rm eV}$
would rule out the quasi-degenerate mass spectrum,
disfavor the spectrum with inverted mass hierarchy
and favor the hierarchical neutrino mass spectrum.

  Using the best fit values of 
$\deltasol$, $\cos2\theta_\odot$ from \cite{SNO2} and
of $\deltaatm$ and $\sin^2\theta$ from \cite{Gonza3nu},
we have found that (Fig. 3, upper panel): 
i) $\meff \ltap 2.0\times 10^{-3}~{\rm eV}$
in the case of neutrino mass spectrum with normal hierarchy,
ii) $2.8\times 10^{-2}~{\rm eV} \ltap \meff \ltap 5.6\times 10^{-2}~{\rm eV}$ 
if the spectrum is with inverted hierarchy, and
iii) $\meff \gtap 0.10~{\rm eV}$ for the quasi-degenerate
mass spectrum. Therefore,
if $\deltaatm$, $\deltasol$ and $\cos2\theta_\odot$
will be determined with a high precision 
($\sim (10-15) \%$ uncertainty)
using the data from
the MINOS, KamLAND and the solar neutrino
experiments and their best fit values 
will not change substantially with respect to those
used in the present analysis~\footnote{The conclusions that follow
practically do not depend on $\sin^2 \theta < 0.05 $.}, 
a measurement of 
$\meff \gtap 0.03~{\rm eV}$ would rule out a hierarchical
neutrino mass spectrum ($m_1 < m_2 \ll m_3$) 
even if there exists a factor of $\sim 6$ (or smaller) 
uncertainty in the value of \meff due to a poor knowledge
of the corresponding nuclear matrix element(s).
An experimental upper limit of $\meff < 0.01~{\rm eV}$    
suffering from the same factor of $\sim 6$ (or smaller) 
uncertainty would rule out the quasi-degenerate 
mass spectrum, while if the uncertainty under discussion is only 
by a factor which is not bigger than $\sim 3.0$, the spectrum with 
inverted hierarchy would be strongly disfavored
(if not ruled out). 

  If the minimal value of  
$\cos2\theta_\odot$ inferred from the solar neutrino 
data, is somewhat smaller than that in eq. (\ref{thLMA}),
the upper bound on \meff in the case 
of neutrino mass spectrum with normal hierarchy 
($\deltasol \equiv \deltadue$, $m_1 \ll 0.02$ eV)
might turn out to be 
larger than the lower bound on \meff 
in the case of spectrum with inverted mass hierarchy 
($\deltasol \equiv \deltatre$, $m_1 \ll 0.02$ eV).
Thus, there will be an overlap
between the regions of allowed values of \meff
in the two cases of neutrino mass spectrum
at $m_1 \ll 0.02 \ {\rm eV}$.
The minimal value of $\cos 2 \theta_\odot$
for which {\it the two regions do not overlap}
is determined by the condition:
\begin{equation}
(\cos 2 \theta_\odot)_{\mbox{}_{\rm MIN}}
= \frac{\sqrt{\deltasolmax} +  2 \sqrt{\deltaatmmax}
(\sin^2 \theta)_{\mbox{}_{\rm MAX}}}
{ 2 \sqrt{\deltaatmmin} + \sqrt{\deltasolmax}}
+ {\cal O} \Big( \frac{\deltasolmax}{4 \deltaatmmin} \Big),
\end{equation}
where we have neglected 
terms of order $(\sin^2 \theta)^2_{\mbox{}_{\rm MAX}}$.
For the values of the neutrino oscillation
parameters used in the present analysis
this ``border'' value turns out to 
be $\cos 2 \theta_\odot \cong 0.25$.

  Let us note that \cite{PPW} if
the \betabeta-decay is not observed,
a measured value of 
$m_{\bar{\nu}_e}$ in 
\hbeta experiments,
$(m_{\bar{\nu}_e})_{exp} \gtap 0.35$ eV,
which is larger than $(m_1)_{max}$,
$(m_{\nu_e})_{exp} > (m_1)_{max}$,
where $(m_1)_{max}$ is determined 
as in the {\it Case 1}
(i.e., from the upper limit
on \meff, $\meff_{\mbox{}_{\rm MIN}} =  \meff_{ \! exp}$,
with $\meff_{\mbox{}_{\rm MIN}}$  given
in eq.~(\ref{minmeffLMA})),
might imply that the massive neutrinos 
are Dirac particles.
If the \betabeta-decay
has been observed and  \meff 
measured, the inequality
$(m_{\bar{\nu}_e})_{exp} > (m_1)_{max}$,
would lead to the conclusion that
there exist contribution(s) to
the \betabeta-decay rate other than 
due to the light Majorana neutrino exchange
which partially cancel the
contribution due to the 
Majorana neutrino exchange.

 A measured value of \meff, 
$( \meff)_{exp} > 0.08 \  \mathrm{eV}$,
and a  measured value of 
$m_{\bar{\nu}_e}$ or an 
upper bound on $m_{\bar{\nu}_e}$,
such that $m_{\bar{\nu}_e} < (m_1)_{min}$,
where $(m_1)_{min}$ is determined by 
the condition
$\meff_{\mbox{}_{\rm MAX}} =  \meff_{ \! exp}$,
with $\meff_{\mbox{}_{\rm MAX}}$  given
by eq.~(\ref{maxmeffLMAinva}), 
would imply that \cite{PPW} 
there are contributions to the
\betabeta-decay rate in addition to the ones
due to the light Majorana neutrino exchange
(see, e.g., \cite{bb0nunmi}), 
which enhance the \betabeta-decay rate.
This would signal the existence of new $\Delta L =2$
processes beyond those induced 
by the light Majorana neutrino exchange
in the case of left-handed charged current  weak 
interaction.

\vspace{-0.3cm}
\section{Conclusions}
\vspace{-0.1cm}

\hskip 1.0truecm Assuming 3-$\nu$ mixing and 
massive Majorana neutrinos,
we have analyzed the implications of the  
results of the 
solar neutrino experiments, including the latest
SNO data, which favor the LMA MSW solution of the 
solar neutrino problem with
$\tan^2\theta_{\odot} < 1$, 
for the predictions of the
effective Majorana mass \meff in 
\betabeta-decay.
Neutrino mass spectra with normal mass hierarchy, 
with inverted hierarchy and of quasi-degenerate type 
are considered. For $\cos 2\theta_{\odot} \geq 0.26$,
which follows (at 99.73\% C.L.)
from the analysis of the solar neutrino data 
performed in \cite{SNO2},
we find significant lower limits on \meff
in the cases of quasi-degenerate and 
inverted hierarchy neutrino mass spectrum,
$\meff \gtap 0.03$ eV and                          
$\meff \gtap 8.5\times 10^{-3}$ eV, respectively. If the 
neutrino mass spectrum is hierarchical (with inverted
hierarchy), the upper limit holds
$\meff  \ltap 8.2 \times 10^{-3}~(8.0 \times 10^{-2})$ eV.
Correspondingly, not only a measured value of 
$\meff \neq 0$, but even an experimental
upper limit on \meff of the order of
${\rm few} \times 10^{-2}$ eV can provide 
information on the type of
the neutrino mass spectrum; it can provide also 
a significant upper limit on the mass of the lightest
neutrino $m_1$. Further reduction of the LMA solution 
region due to data, e.g., from the 
experiments SNO, KamLAND and 
BOREXINO, leading, in particular, to 
an increase (a decreasing) 
of the current lower (upper) bound of
$\cos 2\theta_{\odot}$ can strengthen
further the above conclusions.

   Using the best fit values of 
$\deltasol$, $\cos2\theta_\odot$ from \cite{SNO2} and
of $\deltaatm$ and $\sin^2\theta$ from \cite{Gonza3nu},
we have found that (Fig. 3, upper panel): 
i) $\meff \ltap 2.0\times 10^{-3}~{\rm eV}$
in the case of neutrino mass spectrum with normal hierarchy,
ii) $2.8\times 10^{-2}~{\rm eV} \ltap \meff \ltap 5.6\times 10^{-2}~{\rm eV}$ 
if the spectrum is with inverted hierarchy, and
iii) $\meff \gtap 0.10~{\rm eV}$ for the quasi-degenerate
neutrino mass spectrum. Therefore,
if $\deltaatm$, $\deltasol$ and $\cos2\theta_\odot$
will be determined with a high precision
($\sim (10-15) \%$ uncertainty)
 using the data from
the MINOS, KamLAND and the solar neutrino
experiments and their best fit values 
will not change substantially with respect to those
used in the present analysis, a measurement of 
$\meff \gtap 0.03~{\rm eV}$ would rule out a hierarchical
neutrino mass spectrum ($m_1 < m_2 \ll m_3$) 
even if there exists a factor of $\sim 6$
uncertainty in the value of \meff due to a poor knowledge
of the corresponding nuclear matrix element(s).
An experimental upper limit of $\meff < 0.01~{\rm eV}$    
suffering from the same factor of $\sim 6$ (or smaller) 
uncertainty would rule out the 
quasi-degenerate neutrino mass spectrum,
while if the uncertainty under discussion is 
by a factor not bigger than $\sim 3.0$, the spectrum with 
inverted hierarchy would be strongly disfavored
(if not ruled out). 
 
   Finally, a measured value of $\meff \gtap 0.2$ eV,
which would imply a quasi-degenerate 
neutrino mass spectrum,
combined with data on neutrino masses
from the $^3$H $\beta-$decay 
experiment KATRIN 
(an upper limit or a measured value 
\footnote{ Information on the absolute values
of neutrino masses in the range of interest
might be obtained also from cosmological 
and astrophysical data, see, e.g., ref.~\cite{Weiler2001}.}), 
might allow to establish whether 
the CP-symmetry is violated in the lepton sector.

\vspace{0.3cm}
\leftline{\bf Note Added.} After the 
work on the present study was essentially
completed, few new global analyses of 
the solar neutrino data have appeared
\cite{BargerSNO2,StrumiaSNO2,GoswaSNO2,ConchaSNO3}. 
The results obtained in \cite{BargerSNO2}
do not differ substantially from those derived 
in \cite{SNO2}; in particular,
the (99.73\% C.L.) minimal allowed values of
$\cos 2\theta_{\odot}$
in the LMA solution region
found in \cite{SNO2} and in \cite{BargerSNO2} 
practically coincide.
The best fit values of \deltasol and
$\cos 2\theta_{\odot}$ 
found in \cite{SNO2,BargerSNO2,GoswaSNO2,ConchaSNO3}
also practically coincide, 
with $\cos 2\theta_{\odot}|_{\mbox{}_\mathrm{BF}}$ 
lying in the interval (0.41 - 0.50)
and $\deltasol|_{\mbox{}_\mathrm{BF}} \simeq 5 \times 10^{-5}~{\rm eV^2}$.
The authors of \cite{StrumiaSNO2} find a
similar $\cos 2\theta_{\odot}|_{\mbox{}_\mathrm{BF}}$,
but a somewhat larger 
$\deltasol|_{\mbox{}_\mathrm{BF}} \simeq 7.9\times 10^{-5}~{\rm eV^2}$.
According to \cite{StrumiaSNO2}, \cite{GoswaSNO2} and \cite{ConchaSNO3},
the lower limit $\cos 2\theta_{\odot} > 0.25$
holds approximately at 94\% C.L.,  
90\% C.L. and 81\%~C.L., respectively.
Larger maximal allowed values of
\deltasol than that given in 
eq. (\ref{dmsolLMA}) - of the order of 
$(4 - 5)\times 10^{-4}~{\rm eV^2}$ (99.73\% C.L.),
have been obtained in the analyses
performed in \cite{StrumiaSNO2,GoswaSNO2,ConchaSNO3}.
The authors of \cite{SNO2,BargerSNO2}
used the full SNO data on the day and night event spectra
\cite{SNO2} in their analyses, while the 
authors of \cite{StrumiaSNO2,GoswaSNO2,ConchaSNO3}
did not use at all or used only part of these data.

\begin{figure}
\begin{center}
\epsfig{file=meffnorinvfin.epsi, height=15cm, width=17cm
}
\end{center}
\caption{
The dependence of \meff on $m_1$ 
in the case of the LMA  solution
of the solar neutrino problem \cite{SNO2}
($99.73 \%$~C.L.), for
i) $\deltasol = \Delta m_{21}^2$ (left panels) and
ii) $\deltasol = \Delta m_{32}^2$ (right panels)
and for 
$\sin^2 \theta = 0.05$ (upper panels),
$\sin^2 \theta = 0.01$ (middle panels),
$\sin^2 \theta = 0.005$ (lower panels).
For $\deltasol = \Delta m_{21}^2$,
($\sin^2 \theta = \utre^2$),
the allowed values of \meff 
are constrained to lie
in the case of CP-conservation 
in the medium-grey regions 
{\it a)} 
between the two  thick solid lines if
$\eta_{21} = \eta_{31} = 1$,
{\it b)} 
between the two long-dashed lines and the axes if
$\eta_{21} = - \eta_{31} = 1$,
{\it c)} 
between the dash-dotted lines and the axes
if $\eta_{21} = - \eta_{31} = - 1$,
{\it d)} 
between  the short-dashed lines 
if $\eta_{21} = \eta_{31} = - 1$.
For $\deltasol = \Delta m_{32}^2$, 
($\sin^2 \theta = \uuno$),
the allowed regions
for \meff correspond:
for $|U_{\mathrm{e} 1}|^2 = 0.005$ and
$|U_{\mathrm{e} 1}|^2 = 0.01$ - to 
the medium-grey regions
{\it a})  between the 
solid lines  
if $\eta_{21} = \eta_{31} = \pm 1$,
{\it b})  between the dashed lines
if $\eta_{21} = - \eta_{31} = \pm 1$,
and for $|U_{\mathrm{e} 1}|^2 = 0.05$ - 
to the medium-grey regions
{\it c}) between the 
solid lines  
if $\eta_{21} = \eta_{31} =  1$,
{\it d}) between the 
long-dashed lines  
if $\eta_{21} =  \eta_{31} = -  1$,
{\it e})  between the 
dashed-dotted lines  
if $\eta_{21} = - \eta_{31} =   1$,
{\it f})  between the 
short-dashed lines  
if $\eta_{21} = - \eta_{31} = -  1$,
In the case of CP-violation, the allowed region
for \meff covers all the grey regions. 
Values of \meff in the dark grey regions
signal CP-violation.} 
\label{fig:1}
\end{figure}


\begin{figure}[p]
\begin{center}
\epsfig{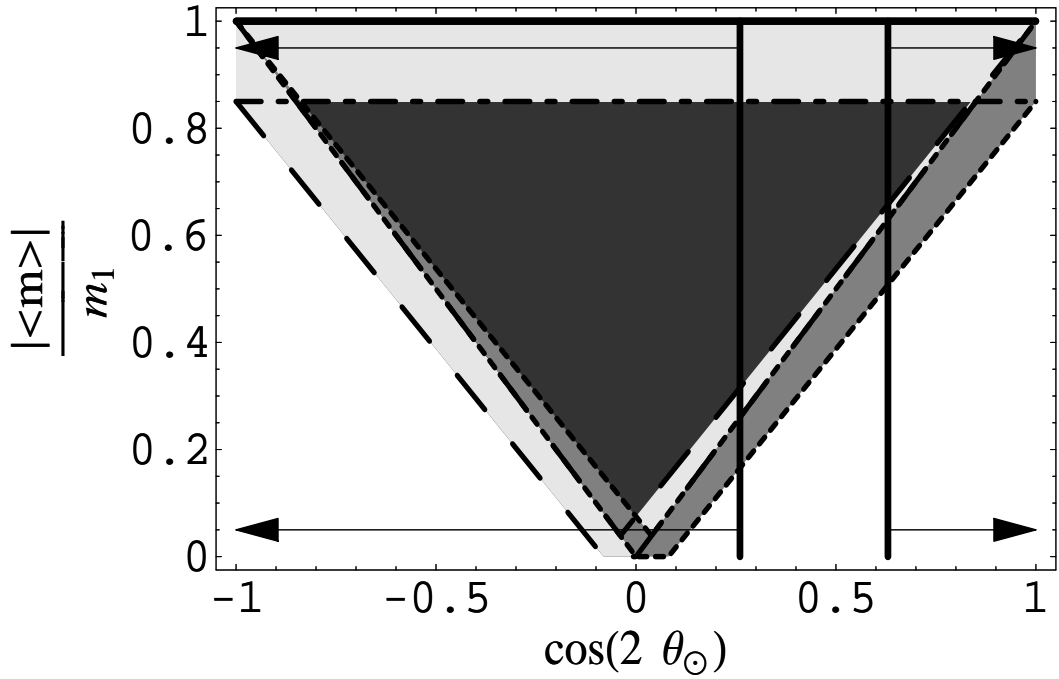}
\end{center}
\caption[gmassdeg03]{
The dependence of $\meff/m_1$ 
on $\cos 2 \theta_\odot$ 
for the quasi-degenerate 
neutrino mass spectrum 
($m_1 > 0.2 \ {\rm eV}$, $m_1 \simeq m_2 \simeq m_3 \simeq m_{\bar{\nu}_e}$).
If CP-invariance holds, the values of 
$\meff/ m_1$ lie:
i) for $\eta_{21}=\eta_{31} = 1$ -
on the line $\meff / m_1 = 1$, 
ii) for $\eta_{21}= - \eta_{31} = 1$ - 
in the region between the  thick horizontal solid and 
dash-dotted lines (in light  grey and 
medium grey colors),
iii) for $\eta_{21}=-\eta_{31} =- 1$ - in 
the light  grey polygon with long-dashed 
and long-dashed-double-dotted line contours
and iv) for $\eta_{21}=\eta_{31} = -  1$ -
in the medium grey 
polygon with the short-dashed  
and long-dashed-double-dotted line contours.
The ``just-CP-violation'' region is denoted 
by dark-grey color. The 
values of $\cos 2 \theta_\odot$ 
between the doubly thick solid  lines
correspond to the lower and upper limits of the
LMA solution regions found 
in ref.~\cite{SNO2}  at $99.73 \%$~C.L. 
} 
\label{fig:2}
\end{figure}

\begin{figure}
\begin{center}
\epsfig{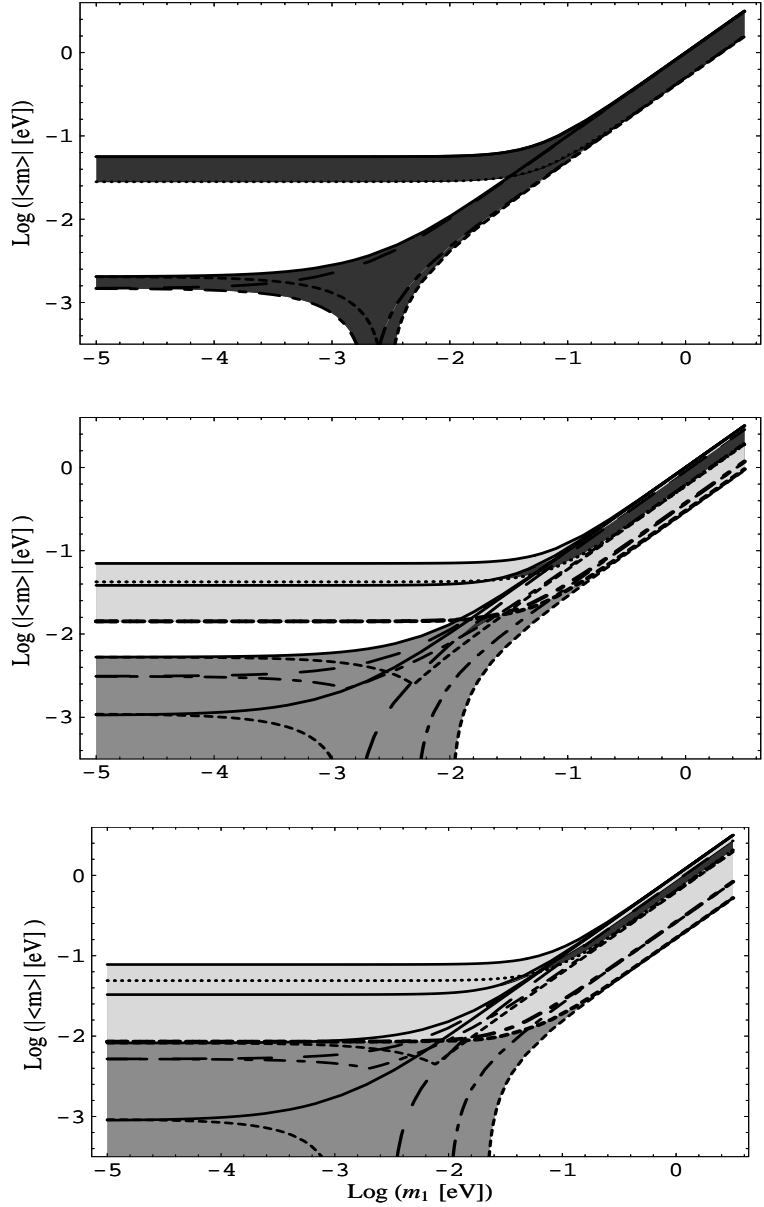}
\end{center}
\vspace{-3mm}
\caption{
The dependence of \meff on $m_1$ 
in the case of the LMA solution, 
for   $\deltasol = \Delta m_{21}^2$ and
$\deltasol = \Delta m_{32}^2$, and 
for the best fit values 
(upper panel) and the $90 \%$~C.L. allowed values 
(middle panel) of the neutrino 
oscillation parameters found in 
refs. \protect\cite{SNO2,Gonza3nu}.
The lower panel is obtained by using the 
$99.73 \%$~C.L. allowed values 
of \deltasol and $\cos2\theta_{\odot}$
from \protect\cite{SNO2} and the 
$99 \%$~C.L. allowed values of \deltaatm and 
$\sin^2\theta$ from \protect\cite{Gonza3nu} 
(the latter article does not include results 
at $99.73 \%$~C.L.).
In the case of CP-conservation,
the allowed values of \meff are constrained 
to lie: for i) $\deltasol = \Delta m_{21}^2$ 
and the middle and lower panels (upper panel) -
in the medium-grey and light-grey regions
{\it a)} between the two lower  thick solid lines 
(on the lower thick solid line) if
$\eta_{21} = \eta_{31} = 1$,
{\it b)} 
between the two  long-dashed lines and the axes 
(on the  long-dashed line) if
$\eta_{21} = - \eta_{31} = 1$,
{\it c)} 
between the two thick  dash-dotted lines and the axes
(on the dash-dotted lines)
if $\eta_{21} = - \eta_{31} = - 1$,
{\it d)}
between the three thick  short-dashed lines and the axes
(on the short-dashed lines)
if $\eta_{21} = \eta_{31} = - 1$;
and for ii) $\deltasol = \Delta m_{32}^2$
and the middle and lower panels (upper panel) -
in the  light-grey regions
{\it a)}
between the two upper  thick solid lines 
(on the upper thick solid line) if
$\eta_{21} = \eta_{31} = \pm 1$,
{\it b)}
between the dotted  and 
the doubly-thick short-dashed lines
(on the dotted line)
if $\eta_{21} = - \eta_{31} = - 1$,
{\it c)}
between the  dotted  and 
the doubly-thick  dash-dotted lines
(on the dotted line)
if $\eta_{21} = - \eta_{31} = + 1$.
In the case of CP-violation, the allowed regions
for \meff cover all the grey regions. 
Values of \meff in the dark grey regions 
signal CP-violation.} 
\label{fig:3}
\end{figure}


\vspace{-0.5cm}

\begin{thebibliography}{99}
\baselineskip 10pt

\bibitem{SNO2} SNO Collaboration,
               Q.R. Ahmad et al., nucl-ex/0204009.

\bibitem{SNO3} SNO Collaboration,
               Q.R. Ahmad et al., nucl-ex/0204008.

\bibitem{SNO1} SNO Collaboration,
               Q.R. Ahmad et al., 
{\em Phys. Rev. Lett.} {\bf 87} (2001) 071301
(nucl-ex/0106015).

\bibitem{SKsol} Super-Kamiokande Collaboration,
                Y. Fukuda {\em et al.}, 
              {\em  Phys.\ Rev.\ Lett. } {\bf 86} (2001) 5651 and 5656.

\bibitem{Cl98}   B.T. Cleveland et al., 
                {\em  Astrophys. J.} {\bf 496} (1998) 505;
                Y.\ Fukuda {\em et al.},
               {\em  Phys.\ Rev.\ Lett.\ } {\bf 77} (1996) 1683;
                V.\ Gavrin, {\em  Nucl. Phys. Proc. Suppl.} {\bf 91} (2001) 36;
                W.\ Hampel {\em et al.},
               {\em  Phys.\ Lett.\ } {\bf B447} (1999) 127;
                M.\ Altmann {\em et al.},
               {\em  Phys.\ Lett.\ } {\bf B490} (2000) 16.


\bibitem{SKatm00} Super-Kamiokande Collaboration,
                  H.\ Sobel et al.,
                 {\em Nucl. Phys. Proc. Suppl.} {\bf 91} (2001) 127.


\bibitem{BGG99}
S. M. Bilenky, C. Giunti and W. Grimus,
{\em Prog. Part. Nucl. Phys.} {\bf 43} (1999) 1.


\bibitem{P99}
           S.T. Petcov, hep-ph/9907216. 


\bibitem{BiPet87} S.M.\ Bilenky and S.T.\ Petcov,
                {\em Rev.\ Mod.\ Phys.} \ {\bf 59} (1987) 67. 


\bibitem{BPP1} S.M. Bilenky, S. Pascoli and S.T. Petcov,
               {\em Phys. Rev.} {\bf D64} (2001) 053010.

\bibitem{PPW} S. Pascoli, S.T. Petcov and L. Wolfenstein,
            {\em Phys. Lett.} {\bf B524} (2002) 319;  
            S. Pascoli and  S.T. Petcov, hep-ph/0111203.


\bibitem{BPont57} B. Pontecorvo, 
                  {\em Zh. Eksp. Teor. Fiz.} {\bf 33}, 549 (1957),
                and {\bf 34}, 247 (1958).
               

\bibitem{MNS62} Z. Maki, M. Nakagawa and S. Sakata, 
{\em Prog. Theor. Phys.} {\bf 28} (1962) 870.


\bibitem{BHP80} S.\ M. \ Bilenky \textit{et al.},
              {\em  Phys.\ Lett.}  {\bf B94} (1980) 495.


\bibitem{Doi81} M.~Doi \textit{et al.},
{\em Phys. Lett.}  \textbf{B102} (1981) 323.


\bibitem{LW81} L. Wolfenstein,
              {\em  Phys.\ Lett.}  {\bf B107} (1981) 77.


\bibitem{BNP84} S.\ M. \ Bilenky, N.\ P.\ Nedelcheva and
S.\ T. \ Petcov, {\em Nucl. Phys.}  \textbf{ B247} (1984) 589;
 B. Kayser, {\em Phys. Rev.} {\bf D30} (1984) 1023.
         

\bibitem{Klap01} H.V. Klapdor-Kleingrothaus \textit{et al.},
                {\em Mod. Phys. Lett.} {\bf 16} (2001) 2409.

\bibitem{FSViss02} F. Feruglio, A. Strumia and F. Vissani, 
hep-ph/0201291.


\bibitem{bb0nu02} C.E. Aalseth \textit{et al.}, hep-ex/0202018.

\bibitem{SPAS94} S. T. Petcov and A. Yu. Smirnov,
                   {\em Phys. Lett.}  \textbf{B322} (1994) 109.

\bibitem{BGKP96} S.M. Bilenky \textit{et al.}, 
{\em Phys. Rev.}  \textbf{D54} (1996) 4432.


\bibitem{BGGKP99} S.M. Bilenky \textit{et al.}, {\em Phys.\ Lett.} 
{\bf B465} (1999) 193.


\bibitem{bbpapers1} V. Barger and K. Whisnant, 
{\em Phys. Lett.}  \textbf{B456} (1999) 194;
H. Minakata and O. Yasuda,
{\em Nucl. Phys.}  \textbf{B523} (1998) 597;
T. Fukuyama \textit{et al.},
{\em Phys. Rev.} \textbf{D57} (1998) 5844 and
hep-ph/0204254;
P. Osland and G. Vigdel,
{\em Phys. Lett.}  \textbf{B520} (2001) 128; 
D. Falcone and F. Tramontano, {\em Phys. Rev.} \textbf{D64} (2001) 077302;
T. Hambye, hep-ph/0201307.


\bibitem{BPP2} S.M. Bilenky, S. Pascoli and S.T. Petcov,
{\em Phys. Rev.} {\bf D64} (2001) 113003. 

\bibitem{bbpapers2}
F. Vissani, {\em JHEP} \textbf{06} (1999) 022;
M. Czakon \textit{et al.}, {\em Phys. Lett.}  \textbf{B465} (1999) 211,
hep-ph/0003161 and {\em Phys. Rev. } {\bf D65} (2002) 053008; 
H.\ V.\ Klapdor-Kleingrothaus, H.\ Pas and A. Yu. Smirnov,
{\em Phys. Rev.} \textbf{D63} (2001) 073005;
H. Minakata and H. Sugiyama, 
{\em Phys. Lett.}  \textbf{B526} (2002) 335;
Z. Xing, {\em  Phys. Rev.} {\bf D65} (2002) 077302;
N. Haba, N. Nakamura and  T. Suzuki, hep-ph/0205141.

\bibitem{KATRIN} A. Osipowicz et al. (KATRIN Project), hep-ex/0109033.

\bibitem{bbpapers3} 
W. Rodejohann, {\em Nucl. Phys.} {\bf  B597} (2001) 110, and 
hep-ph/0203214.


\bibitem{76Ge00}
        H.\ V.\ Klapdor-Kleingrothaus \textit{et al.},
{\em Nucl. Phys. Proc. Suppl.} {\bf 100} (2001) 309.


\bibitem{IGEX00} C.E. Aalseth, F.T. Avignone III et al.,
{\em Physics of Atomic Nuclei} {\bf 63} (2000) 1225.


\bibitem{NEMO3} X. Sarazin \textit{et al.} (NEMO3 Collaboration),
     hep-ex/0006031.

\bibitem{CUORE}  E. Fiorini, {\em Phys. Rep.} {\bf 307} (1998) 309.
          

\bibitem{GENIUS} H.\ V.\ Klapdor-Kleingrothaus \textit{et al.},
          {\em J. Phys.} G \textbf{24} (1998) 483.

\bibitem{EXO}  M.\ Danilov \textit{et al.},
	  {\em Phys.\ Lett.} {\bf B480} (2000) 12.

\bibitem{Maj} L. De Braeckeleer et al. (Majorana Project),
          {\em Proceedings of the Carolina Conference on Neutrino Physics}, 
           Columbia (SC), USA, March 2000.

\bibitem{MOON}
H. Ejiri {\em et al.}, {\em Phys. Rev. Lett.}
                         {\bf 85} (2000) 2917.


\bibitem{MoscowH3} V.\ Lobashev \textit{et al.},  
                 {\em Nucl. Phys. Proc. Suppl.} {\bf 91 }(2001) 280.


\bibitem{Mainz} Ch. Weinheimer \textit{et al.},
talk at the Int. Conf. on Neutrino Physics and 
Astrophysics ``Neutrino'02'', May 25 - 30, 2002, Munich, Germany.


\bibitem{CHOOZ} M.\ Appolonio \textit{et al.}, 
                 {\em Phys. Lett. }{\bf B466} (1999) 415.


\bibitem{PaloV} 
F. Boehm et al., 
{\em Phys.\ Rev.\ Lett.\ }  {\bf 84} (2000)  3764 and
{\em Phys. Rev.} {\bf D62} (2000) 072002.



\bibitem{Gonza3nu} C. Gonzalez-Garcia \textit{et al.},
                   {\em Phys. Rev.} {\bf D63} (2001) 033005.


\bibitem{MINOS} D. Michael (MINOS Collaboration),
talk at the Int. Conf. on Neutrino Physics and 
Astrophysics ``Neutrino'02'', May 25 - 30, 2002, Munich, Germany.
  		
\bibitem{KamLAND}  J. Shirai (KamLAND Collaboration), talk at the
Int. Conf. on Neutrino Physics and Astrophysics ``Neutrino'02'',
May 25 - 30, 2002, Munich, Germany.


\bibitem{Carlos01} A. de Gouv\^ea and C. Pe\~na-Garay, hep-ph/0107186. 


\bibitem{bb0nunmi} K.S. Babu and R.N. Mohapatra,
{\em Phys. Rev. Lett.} {\bf  75} (1995) 2276;
H. Pas et al., {\em Phys. Lett. } {\bf   B398} (1997) 311; {\it ibid.} 
{\bf B459} (1999) 450.


\bibitem{Weiler2001} T. J. Weiler, 
                     {\em Astropart. Phys. } {\bf 11}
                     (1999) 303; 
H. Pas and T. J. Weiler,  
       {\em Phys. Rev. } {\bf  D63} (2001) 113015; 
        Z. Fodor {\it et al.},
     {\em  Phys. Rev. Lett. } {\bf  88} (2002) 171101.


\bibitem{BargerSNO2} V. Barger et al., hep-ph/0204253, 
version 2 of 29 of April, 2002.

\bibitem{StrumiaSNO2} P. Creminelli, G. Signorelli and A. Strumia,
hep-ph/0102234, version 3 of 22 of April, 2002.

\bibitem{GoswaSNO2} A. Bandyopadhyay \textit{et al.}, hep-ph/0204286.

\bibitem{ConchaSNO3} J.N. Bachall, M.C. Gonzalez-Garcia and C. Pe\~{n}a-Garay,
hep-ph/0204314.

\end{thebibliography}
\end{document}